\documentstyle[12pt,aaspp4]{article}
\newcommand{\be}{\begin{equation}}
\newcommand{\ee}{\end{equation}}
\newcommand{\ba}{\begin{eqnarray}}
\newcommand{\ea}{\end{eqnarray}}
\begin{document}
\title{Reconnection in the ISM} 
\author{Ethan T. Vishniac\altaffilmark{1} and Alex Lazarian\altaffilmark{2}}
\altaffilmark{1}{Department of Astronomy, University of Texas, Austin TX
78712}
\altaffilmark{2}{Princeton University Observatory, Princeton, NJ 08544}
\begin{abstract}

We discuss the role of ambipolar diffusion for simple reconnection
in a partially ionized gas, following the reconnection geometry
of Parker and Sweet.
When the recombination time is short the mobility and reconnection of 
the magnetic field is substantially enhanced as matter escapes
from the reconnection region via ambipolar diffusion.
Our analysis shows that
in the interstellar medium it is 
the recombination rate that usually limits the
rate of reconnection. Consequently, the typical
reconnection velocity in interstellar medium
is $\sim(\eta/\tau_{recomb})^{1/2}(1+x^{-1}\beta^{-1})$,
where $\eta$ is the ohmic resistivity, $x^{-1}$ is the ionization
fraction, and $\beta$ is the ratio of gas pressure to magnetic
pressure.  We show that heating effects can reduce this speed
by increasing the recombination time and raising
the local ion pressure. In the colder parts of the ISM,
when temperatures are $\sim 10^2$K or less, we obtain a significant
enhancement over the usual Sweet-Parker rate, but only in dense molecular 
clouds will the reconnection velocity exceed $10^{-3}$ times the
Alfv\'en speed.  The ratio of the ion orbital radius to
the reconnection layer thickness is 
typically a few percent,
except in dense molecular clouds where it can approach unity.
We briefly discuss prospects for obtaining much faster reconnection
speeds in astrophysical plasmas.
\end{abstract}


\section{Introduction}

Understanding the mobility and reconnection of magnetic fields in a conducting
medium is critical to understanding the evolution and origin of
large scale magnetic fields in astrophysical objects such as
stars, galaxies, and accretion disks.  Standard
dynamo theory relies on the turbulent transport of magnetic 
flux to move the field lines and, implicitly, to change the
large scale topology of the magnetic field.  However, the substitution
of turbulent transport for microscopic diffusion is difficult to
justify on theoretical grounds.  Normally we would expect flux
freezing to be an excellent approximation to the motion of the
magnetic field in a highly conducting fluid.  Getting field lines
to pass through one another, or rearrange their connections, must
ultimately involve ohmic diffusion.  Moreover, as the field lines
are stretched, they will strengthen and exert forces that will prevent
further stretching or deformation.  Recently several different
authors (e.g. Cattaneo and Vainshtein 1991, Gruzinov and Diamond
1994) have argued that this raises an insuperable obstacle to
turbulent transport, and standard mean-field dynamo theory, in
highly conducting fluids.  One of us (\cite{V95}) has proposed
that this problem may be solved through the formation of intense
flux tubes, but this assumes that reconnection is rapid enough
to allow the formation of such structures in a small number of
dynamical time scales and that the plasma has a high $\beta$
so that the magnetic field can be distributed intermittently.

Estimates of reconnection speeds based on a simplified
geometry (\cite{S58}, \cite{P57}) give 
\be
V_{rec}\approx \left({v_A\eta\over L_x}\right)^{1/2}=v_A Re_B^{-1/2}
\label{eq:sp1}
\ee
where $L_x$ is the typical structure scale, $v_A$ is the Alfv\'en
speed, $\eta$ is the ohmic diffusion constant, and $Re_{B}\equiv (v_AL_x/\eta)$ 
is the magnetic Reynolds number.  In general, we expect that $v_A$ is
comparable to the local turbulent velocity, so that this speed
will be many orders of magnetic slower than typical fluid velocities
in astrophysical objects, where $Re_B>>1$.  However, we note
that observations of magnetic fields in the solar chromosphere and
corona (cf. Dere 1996, Innes et al. 1997) suggest that reconnection can occur
at speeds $\sim 0.1 v_A$.  It is not clear what conditions are
necessary for such rapid reconnection or what mechanism is operating
that allows it.

In some ways, these problems are especially severe when we consider
the evolution of the galactic magnetic field (\cite{KA92}).  
The huge scales
involved limit the number of dynamical time scales available for
generating a large scale field and strong theoretical objections
have been raised against the possibility of a primordial field
strong enough to eliminate the necessity for a galactic dynamo.
In addition, the magnetic Reynolds number in the galactic disk
is of order $10^{20}$, so that the Sweet-Parker reconnection
speed is negligiblely small.  

On the other hand, the interstellar medium is only partially ionized,
and in dense, cool clouds the ionization fraction is very small.
This raises the possibility that magnetic field lines which move
with the ion velocity may nevertheless be capable of a high degree of
mobility relative to the bulk of the gas.  Ultimately, the speed of 
magnetic reconnection is limited by the width of the current sheet
dividing magnetic fields of sharply different orientation. 
Although importance of ambipolar diffusion has been long
stressed (see Mestel 1985), the results obtained so far are
contradictory.  Recent one dimensional numerical 
work by Brandenburg and Zweibel (1995) and Zweibel and Brandenburg
(1997)
suggests that ambipolar diffusion can give rise to narrow current
sheets, whose widths are ultimately determined by the size of
particle orbits in the plasma.  This in turn suggests that reconnection in the 
neutral parts of the interstellar 
medium may play a critical role in the galactic dynamo (cf. Subramanian 1998).
At the same time analytical studies of reconnection in the presence of
ambipolar diffusion by Dorman (1996), also based on
a one dimensional model, led to a different conclusion.

These ambiguities motivate our current study.
In this paper we reexamine the role of ambipolar diffusion
in Sweet-Parker reconnection, including the effects of ion pressure
and finite recombination rates for the ions, and allowing for the
{\it transverse} loss of neutral particles during reconnection. 
We calculate reconnection
velocities and show that for most astrophysically important cases
reconnection is limited by the recombination rate. We find
a substantial enhancement of the reconnection rates even in media with
low rates of recombination. This enhancement is not enough, however,
to account for magnetic flux tube formation in the ISM (as
discussed in Lazarian \& Vishniac (1996) and Subramanian (1998)).
At the same time the rates obtained here can explain topological changes 
that accompany stellar formation, that is, the disconnecting of the magnetic
field of a collapsing cloud from the interstellar magnetic field.

In section II of this paper we will briefly review the physical
basis for the Sweet-Parker rate, and for the enhancement due to
ambipolar diffusion.  In section III we estimate the effects of
ion pressure and give a simple mathematical model supporting our
claims.  In section IV we calculate rates of reconnection 
for different phases of the interstellar medium. 
{}Finally, in section V we discuss the implications and limitations
of our work, and summarize our conclusions.

\section{Sweet-Parker Reconnection and Ambipolar Diffusion}

\subsection{The Case of the Imperfect, Ionized Gas}

We start by reviewing the physical basis for the Sweet-Parker
reconnection rate.  This is well known material, but including
it here will make subsequent arguments concerning ambipolar diffusion and
recombination clearer.
The simple geometry that forms the basis of the Sweet-Parker
reconnection rate consists of two regions of opposite magnetic 
polarity facing one another.  The $\hat y$ axis is perpendicular
to the field lines and gas is approaches
the midplane with a velocity $-u_y$ for $y>0$ and $u_y$ for
$y<0$.  The magnetic field lines curve apart a distance
$\pm L_x$ along the $\hat x$ axis.  As the field lines
curve apart the gas streams to the right and left with 
a velocity $u_x$.  The magnetic field
is zero along the $\hat x$ axis, but rises to $\pm B$ a
distance $L_y$ on either side.  This geometry is illustrated
in figure 1, although the arrows for gas flow are appropriate
for the case of ambipolar diffusion.  When the gas is completely
ionized the flow along the $\hat x$ axis is confined to the layer
of width $\Delta$. 

The evident neglect of the
existence of a third dimension is the major weakness of
this picture.  However, the tension of the magnetic field lines, and
their consequent resistance to bending, makes the role of
this neglected dimension quite complicated.  Here we will simply assume
that this geometry provides a useful approach to the problem
of reconnection, and defer consideration of more complicated
geometries to a subsequent paper.  It is important to keep
in mind that in three dimensions the magnetic field along the reconnection
surface does not actually vanish, but simply reduces to a 
component in the $\hat z$ direction which is common to both
magnetized  regions.  This does not effect our arguments in
any way, but implies that one should resist the temptation
to think of the plasma in the reconnection zone as essentially
unmagnetized.  

Conservation arguments can be used to estimate $u_x$ and constrain
the reconnection geometry.  First, we note that the pressure
along a line parallel to the $\hat y$ axis and passing through the origin 
is approximately constant, i.e.
\be
P+{B^2\over 8\pi}\approx\hbox{constant}.
\ee
At the midplane the magnetic pressure
vanishes.  Following the fluid to the right or left, one 
comes (in a distance $\sim L_x$) to a region unaffected by
the magnetic pressure.  This implies a pressure excess in the
reconnection region of $B^2/8\pi$.  This pressure excess is
sometimes derived by considering the dissipation of magnetic
energy in the reconnection layer, but this gives rise to the
mistaken notion that an efficiently radiating 
reconnection layer will not have a gas pressure excess.
{}From Bernoulli's theorem
\be
u_x^2+{P\over \rho}\approx\hbox{constant}~~~,
\ee
which implies that if $u_x=0$ at the origin then at $x=\pm L_x$
\be
u_x=\left({\Delta P\over\rho}\right)^{1/2}\approx 
{B\over \sqrt{8\pi\rho}}={v_A\over 2^{1/2}}~~~,
\ee
where $v_A$ is the Alfv\'en velocity of the magnetized regions.

The mass influx into the reconnection region is just $2\rho L_x u_y$,
while the mass efflux is $2\rho L_x u_y=2\rho L_y u_x$, so from
conservation of mass we have
\be
u_y={L_y\over L_x} v_A~~~.
\label{cons}
\ee

We note that the ions need not trace the
movement of the magnetic flux in an imperfect fluid.  The
magnetic field current is
\be
{\bf J}_B={\bf v}_{i}\bigotimes{\bf B} -\eta {\bf\nabla}{\bf B}~~~,
\label{eq:curr1}
\ee
where the subscript $i$ refers to the ions.
Given the geometry used here, the effective velocity of the
magnetic field can deviate from $u_y$ by an
amount $\eta/L_y$.  To put it another way, the magnetic field is
no longer flux frozen if $L_y$ is sufficiently small and regions of
opposite polarity can annihilate much faster than matter can be
dragged into the reconnection region.  This acts as a kind of
regulator for the size of the reconnection region.  If it is too
large, then reconnection slows to a crawl and whatever external
forces are pushing the magnetized regions together will continue
to do so\footnote{The role of the external forcing is, in fact, more complex.
It can squeeze the material from the reconnection zone and thus increase
the local Alfv\'en velocity.}.  If it is too small, then reconnection runs away and broadens
the neutral zone.  We can therefore take $\eta\approx L_y u_y$ and obtain
\be
u_y\sim\left({\eta v_A\over L_x}\right)^{1/2}~~~.
\label{eq:ps}
\ee
This is the Sweet-Parker reconnection rate given in equation (\ref{eq:sp1}).

One important aspect of this model is the existence of a pressure deficit,
of order the magnetic pressure, at a distance $\pm L_x$ along the $\hat x$
axis.  In a turbulent medium with equipartition between the magnetic and
turbulent energies such pressure excesses will come and go on a time scale
not much longer than $L_x/v_A$, so that this stationary model can be no more
than a very approximate guide to the structure of the reconnection layer. 
However, the existence of magnetic tension guarantees that if the magnetic
field reaches a locally persistent equilibrium, then we can naturally expect
such pressure fluctuations to last as long as the magnetic structure itself.
In other words, the tension in magnetic `knots' is sufficient to guarantee
a downstream pressure deficit of order $B^2/8\pi$.  One might suppose
that the accumulation of ejected ions will erase this pressure deficit, but
as reconnection proceeds the reconnected magnetic field lines will be 
pulled away from the reconnection region, bearing with them the ejected ions.
It is also true, but not obvious, 
that ejected neutrals can usually be ignored in this model, a point we will 
return to later.

\subsection{The Case of the Imperfect, Partially Ionized Gas}

When the fluid is partly neutral the charged and neutral
particles will no longer move together, and ${\bf v}_{i}$
is no longer the same as the bulk velocity of the fluid. 
We can estimate ${\bf v}_{i}-{\bf v}_{n}$ by balancing
the pressure exerted on the ions with the collisional
drag due to collisions with the neutrals.
We get
\be
{{\bf\nabla}B^2\over 8\pi}+
c_i^2{\bf\nabla}\rho_i=\rho_{i}{{\bf v}_{n}-{\bf v}_{i}\over t_{i,n}}
=\rho_{n}{{\bf v}_{n}-{\bf v}_{i}\over t_{n,i}}=-c_n^2{\bf\nabla\rho_n}~~~,
\ee
where $t_{i,n}$ and $t_{n,i}$ are the collision times for ions impacting on
neutrals and vice versa, respectively, and $c_i$ and $c_n$ are the ion and
neutral sound speeds.  
If we neglect the ion pressure gradient (more on this later) we can write
the neutral particle velocity as
\be
{\bf v}_n={\bf v}_i+{{\bf\nabla}B^2\over 8\pi\rho_n}t_{n,i}~~~,
\ee
or
\be
{\bf v}_i={\bf v}-{{\bf\nabla}B^2\over 8\pi\rho}t_{n,i}~~~,
\ee
where $\rho$ and ${\bf v}$ are the total density and bulk velocity
of the gas.  This implicitly assumes
that the neutrals are moving relatively slowly, and dominate the fluid mix,
so that the neutral pressure gradient can balance the magnetic pressure
when the ion pressure gradient fails to do so. 
We can use this expression in equation (\ref{eq:curr1})
to write the magnetic field current as
\be
{\bf J}_B={\bf v}\bigotimes {\bf B} -(\eta+v_A^2t_{n,i}){\bf\nabla}{\bf B}~~~,
\label{eq:curr2}
\ee
where $v_A$ is the Alfv\'en velocity relative to the total gas density.
{}From here on we will define 
$\eta_{ambi}\equiv v_A^2t_{n,i}$.  Typically $v_A\sim 10^5$ cm/sec.
Since $t_{n,i}\sim 4.8\times10^8/n_i$~seconds
at low temperatures (\cite{D58}), and $\eta$ is of order $10^9$ cm$^2$/sec,
$\eta_{ambi}$ will be many orders of
magnitude larger than $\eta$ in cool, low density regions in the interstellar 
medium.

Equation (\ref{eq:curr2}) seems to imply that we can use $\eta_{ambi}\gg \eta$
in place of the usual expression for Ohmic diffusion.  The neutral gas will
stream outward in the $\hat x$ direction with a velocity $\sim v_A$, driven
by the pressure gradient created by collisions with the in-flowing ions.   
In reality this derivation includes assumptions which impose
severe constraints on the substitution of $\eta_{ambi}$ for $\eta$.  
The most important is that
we have assumed that the ion pressure gradient is
negligible compared to the magnetic pressure gradient.
However, if the ions move inward at a steadily increasing speed
towards the reconnection surface then for $y\ll L_y$ the only
escape route for the accumulating particles is to join the 
opposing flow of neutrals. This
can only be true if the recombination time is short, so that
ions and neutrals change identity easily. This expression is commonly
used in calculating the ambipolar diffusion time for molecular
clouds, in which case one only requires the recombination time to
be no greater than the time it takes the magnetic field to drift
out of a cloud.  

Here we are concerned with the formation of
a reconnection layer, in which the relevant time scales
are much shorter.  We begin by assuming rapid recombination. 
In that case we can see that 
the width of the reconnection region, $L_y$
is set by the matter diffusion, as above, by equating the drift
velocity to $u_y$.  This implies
\be
u_y\sim\left({v_A^3t_{n,i}\over L_x}\right)^{1/2}
=\left({v_A\eta_{ambi}\over L_x}\right)^{1/2}~~~,
\label{uy}
\ee
when $\eta_{ambi}>\eta$. Later on we will refer to this
expression for $u_y$ 
as $V_{max}$ to emphasize that this is the maximal velocity
of reconnection that is obtainable through ambipolar diffusion.
By the same reasoning we have
\be
L_y\sim \left({\eta_{ambi} L_x\over v_A}\right)^{1/2}~~~.
\label{Ly1}
\ee

In obtaining expressions (\ref{uy}) and (\ref{Ly1}) we assumed, first,
that the outflow from the region $L_y$ happens with a velocity $v_A$ and,
second, that on leaving the reconnection layer neutrals diffuse
{\it slowly} out  while moving together with ions over the distance
comparable with $L_x$. The former assumption depends on the
pressure differential around the reconnection region, which
is assumed as part of our basic geometry, and the dynamics
of the outflow, which we check below by
studying the structure of the ambipolar diffusion layer. 
The latter assumption constrains the geometries of the reconnection 
layers considered.

It's important to note that the actual process of 
reconnection takes place only in a narrow zone where $\eta>\eta_{ambi}$.
However, unlike the usual Sweet-Parker result, the width of this
inner zone does not determine the reconnection speed.  Instead its
width automatically adjusts itself to match the reconnection speed
determined by ambipolar diffusion.  Ambipolar diffusion removes
matter from the reconnection zone and enhances the reconnection
velocity. 

The existence of this broad outflow raises the possibility of
a problem in our model.  Since the gas outflow is not confined to the 
layer where the magnetic
field is actually recombining, we cannot assume that the
ejected gas will be removed from the downstream
flow by the relaxation of the reconnected field lines.
Instead, the neutral component of the gas has to diffuse
into the wedge of reconnected field lines by overcoming
the ion-neutral drag.  Since the downstream pressure
deficit will help push the neutrals in this direction,
this diffusion will occur at a rate $\sim \eta_{ambi}/L_y^2$.
In the mean time the transverse motion of the neutrals will be
approximately along the field lines, since the ions and neutrals
are strongly coupled through collisions and move together
at a speed $\sim v_A$.  Consequently, the gas will move a distance
$v_A L_y^2/\eta_{ambi}\sim L_x$  while diffusing onto the
reconnected field lines.  It is a basic condition of this model
that the reconnection region is embedded in some larger flow, allowing
room for the relaxation of field lines and the escape of ejected ions.
Since this distance is no larger than the size of the reconnection
region it is a short enough that we can assume that the
accumulation of ejected gas does not pose a problem for the continued flow
of gas from the reconnection region.

We can estimate the width of the resistive zone by considering the motion
of the field lines for $y<L_y$.  Within this zone the magnetic field lines 
speed up as the pressure gradient steepens.  Since the bulk velocity remains 
comparable to $V_{rec}$, while the magnetic field strength plummets, 
the magnetic flux is carried by the second term on
the LHS of equation (\ref{eq:curr2}).  Consequently, we have
\be
V_{rec}B_\infty\approx \eta_{ambi}\partial_y B~~~,
\ee
for $y\ll L_y$ and
where $B_\infty$ is the magnetic field strength far outside 
the reconnection layer.  This has the solution
\be
B=B_\infty\left({3 V_{rec}\over\eta_{ambi,\infty}}y + C\right)^{1/3}~~~,
\label{eq:bamp}
\ee
where $\eta_{ambi,\infty}$ is the 
ambipolar diffusion coefficient far from the reconnection zone and 
$C$ is a constant equal to $B(y=0_+)^3/B_\infty^3$.  (The magnetic field
strength near $y=0$, but outside the layer where reconnection actually
occurs, is defined as $B(0_+)$.) 

The ion velocity, which is also the inward speed of the magnetic field lines, is
given by
\be
v_i=\eta_{ambi}\partial_y\ln B=V_{rec}\left({3V_{rec}\over\eta_{ambi,\infty}}y+
C\right)^{-1/3}~~~.
\label{eq:vion}
\ee
If $C=0$ then this expression diverges near $y=0$.  One resolution of this problem
is to note that $v_i$ is limited by the local value of $v_A/x^{1/2}$, where $x$
is the ionization fraction of the gas.  As $v_i$ approaches this value
the assumption that the acceleration term is negligible is violated and
we have $v_i\rightarrow v_A/x^{1/2}$.  This implies a limit on $C$, which can be used
as an estimate for $B(y=0_+)$.  Using equations (\ref{eq:bamp}) and (\ref{eq:vion}) 
we find
\be
B(0_+)\sim B_\infty \left({x\eta_{ambi,\infty}\over L_x v_{A,\infty}}
\right)^{1/4}= 
B_\infty \left({x v_{A,\infty}t_{n,i}\over L_x}\right)^{1/4}~~~.
\ee
This expression will be relevant provided that $\eta$ remains less than $\eta_{ambi}$
as the magnetic field approaches this asymptotic limit.  More precisely, this
is relevant when
\be
\eta<\eta_{ambi,\infty}
\left({x \eta_{ambi,\infty}\over L_x v_{A,\infty}}\right)^{1/2}~~~.
\label{eq:fastcrit}
\ee
In this case the asymptotic value of $v_i$ is
\be
v_i(0_+)=v_{A,\infty}\left({v_{A,\infty} t_{n,i}\over x L_x}\right)^{1/4}~~~,
\ee
and the width of the resistive reconnection region is
\be
\Delta\approx {\eta\over v_i(0_+)}\approx 
\left({L_x\eta\over v_{A,\infty}}\right)^{1/2}x^{1/4}
\left({\eta\over\eta_{ambi,\infty}}\right)^{1/4}
\left({\eta\over L_xv_{A,\infty}}\right)^{1/4}~~~.
\label{eq:w1}
\ee

In either case, we expect that $B(0+)\ll B_{\infty}$ and consequently
that magnetic pressure scales as y$^{2/3}$ throughout the outer
layers of the reconnection region. This implies that the
overpressure in the layer $L_y$ scales as $B^2_{\infty}(1-(y/L_y)^{2/3})$
and that the outflow velocity will be of the order of $v_A$.
This justifies {\it a posteriori} our assumption of
rapid outflow in this case.

When the condition expressed in equation (\ref{eq:fastcrit}) is not
satisfied $v_i$ stays below the ion Alfv\'en speed at all times.
In this case we can take $C=0$ and determine the width of the
resistive reconnection region from the condition that
\be
{\eta\over\Delta}\approx |v_i(y=\pm\Delta)|~~~.
\ee
{}From equation (\ref{eq:vion}) this implies that
\be
\Delta\approx \left({L_x\eta\over v_{A,\infty}}\right)^{1/2}
{\eta\over\eta_{ambi,\infty}}~~~.
\label{eq:w2}
\ee

We see from equations (\ref{eq:w1}) and (\ref{eq:w2}) that 
the width of the resistive reconnection region is much smaller 
in this case than when ambipolar diffusion is negligible. 
This is expected, since a narrow reconnection region is
necessary for a fast reconnection speed, but it also raises
the question of local heating.  The bulk of the
magnetic energy is dissipated through expansion as the
ions speed up, there is an irreducible minimum which is
annihilated inside the reconnection region proper.  The
volume heating rate will be
\be
\dot{\cal E}\approx B(y=0_+)^2{\eta\over\Delta^2}~~~,
\label{dE}
\ee
but since $\Delta=\eta/v_i$ this is
\be
\dot{\cal E}\approx B(y=0_+)^2{v^2_i\over\eta}=B_\infty^2 {V_{rec}^2\over\eta}
=B_\infty^2 {v_{A,\infty}\over L_x}{\eta_{ambi,\infty}\over\eta}~~~.
\ee

We conclude that the local heating rate is significantly enhanced
in the center of the reconnection region, in the resistive layer, and that this may
introduce difficulties for a model which relies on an abundance
of neutral particles to carry mass away from the reconnection region.
We will not attempt to calculate the consequences of this heating
here.  It is probably not the most severe constraint on ambipolar
diffusion in the interstellar medium.  Instead, we suggest that
the long recombination time for ions, coupled to the effects
of ion pressure, poses a much greater
problem.  We examine this point in the next section of this paper.

\section{The Role of Ion Pressure}

The critical assumption in the preceding section is that it is
reasonable to assume that the ions and neutral particles maintain
their usual ratio within the reconnection layer.  We can see why this
important by restricting our attention to the resistive layer.  The
net inward flux of ions will be $>V_{rec}n_{i,\infty}$.  This constitutes a
lower limit, since within the whole reconnection region of width $L_y$
the magnetic field lines speed up, so that their density drops, and
the same effect will depress $n_i$ below its equilibrium value.
A uniform neutral distribution will therefore {\it add} incoming
ions to this flow. 

The rate at which ions leave the resistive
region through recombination is
\be
n_i(0){\Delta\over \tau_{recomb}(0)} =
n_i(0){\eta\over v_i(0_+)\tau_{recomb}(0)}> n_{i,\infty} V_{rec}~~~.
\ee

This assumes that the loss of ions through their expulsion in the
$\hat x$ direction is negligible, but if that process dominates then
we will recover the usual Sweet-Parker result.  When the recombination
rate is small, ions will accumulate in the resistive layer creating
a strong ion pressure gradient, and the reconnection process will
be limited by the rate at which ions can recombine with electrons 
within the resistive layer.

We can understand the role of the ion pressure by equating the ion 
density flux with the recombination 
losses in the resistive layer. We have
\be
v_i(0_+)\rho_i(0_+)={2\rho_i(0)\Delta \over\tau_{recomb}(0)}~~~.
\label{eq:ion1}
\ee
This implies that
\be
v_i(0_+)=
\left({2\eta(0)\over\tau_{recomb}(0)}\right)^{1/2}
\left({T(0_+)\over T(0)}+{V_{rec}\over v_i(0_+)}
{B^2_\infty\over8\pi\rho_{i,\infty}c_i^2(0)}\right)^{1/2}~~~,
\label{eq:vrec1}
\ee
where we have invoked pressure balance for the ion density in the
resistive layer, and assumed that $B/\rho_i$ is a constant
outside of the resistive layer.  

The recombination rate is $\alpha n_e$, where $n_e\approx n_i$ and
$\alpha$ is the recombination coefficient (see Spitzer 1978, p. 107),
which scales approximately as $2\times 10^{-11}\phi(T)/T^{1/2}$ cm$^3$ s$^{-1}$
for a hydrogen plasma, where $\phi$ is a slowly varying function of 
temperature.  
In an ionized plasma the resistivity is $\approx 2.65\times10^{12}T^{-3/2}$, 
and this expression should remain valid as long as the scattering
length for electrons is determined by inelastic collisions with ions. 
In a largely neutral gas the exact
dependence on density and temperature can be fairly complicated.
However, for our purposes a crude estimate is all that is required
and in any case the density of ions in the resistive layer is greatly 
enhanced in order to balance the magnetic pressure in the surrounding plasma.
Using these results we can
rewrite equation (\ref{eq:vrec1}) as
\be
v_i(0_+)=
\left({2\eta(0_+)\over\tau_{recomb}(0_+)}\right)^{1/2}
\left({T(0_+)\over T(0)}\right)^{3/2}
\left(1+{V_{rec}\over v_i(0_+)}
{B^2_\infty\over8\pi\rho_{i,\infty}c_i^2(0_+)}\right)~~~,
\label{eq:vrec2}
\ee

Equation (\ref{eq:vrec2}) is only 
an upper limit on $V_{rec}$, which in principle could be much
lower if there exists an outer layer dominated by
ion-neutral drag.  However, we can show that this is unlikely
to be the case.  If such a layer exists then 
we can see from equation
(\ref{eq:vion}) that the ion velocity in this layer
becomes 
\be
v_i=V_{rec}\left({3V_{rec}\over\eta_{ambi,\infty}}y+
\left({V_{rec}\over v_i(0_+)}\right)^3\right)^{-1/3}~~~.
\label{eq:vion1}
\ee
This in turn implies an outer layer width of
\be
L_y\sim {\eta_{ambi,\infty} V_{rec}^2\over v_i(0_+)^3}~~~,
\ee
and a reconnection velocity of
\be
V_{rec}=\left({v_i(0_+)^2L_x\over v_A \eta_{ambi,\infty}}\right)v_i(0_+)~~~.
\ee
This could represent an enormous reduction of $V_{rec}$
from its upper limit of $v_i(0_+)$, but this outer layer will
not form unless the cumulative drag from the neutrals in this
layer is large enough to lead to a significant reduction in the
local magnetic field strength.  Otherwise the outer layer will
collapse and $v_i(0_+)\rightarrow V_{rec}$.  

At large distances the ions and neutrals move together with a 
velocity $V_{rec}$ towards the reconnection zone.  The accumulation
of neutrals will lead them to decelerate and accumulate in a
layer of width $L_n$.  If we assume that this layer is much thicker
than the actual zone of reconnection, then the relative ion-neutral
velocity in this layer will be $\sim V_{rec}$ and the net drag force
will be 
\be
\rho_n V_{rec}{L_n\over t_{n,i}}\sim \delta P_n,
\label{v1}
\ee
where we have assumed that the drag force is balanced by the excess
neutral pressure.  The neutrals will be expelled laterally
at a speed $v_{x,n}$ given by
\be
v_{x,n}^2\sim{\delta P_n\over\rho_n}.
\label{v2}
\ee
Using Eqs. (\ref{v1}), (\ref{v2}) and conservation
of mass we conclude that 
\be
v_{x,n}\sim \left({V_{rec}^2 L_x\over t_{n,i}}\right)^{1/3},
\label{eq:vxn}
\ee
and
\be
L_n\sim L_x\left({V_{rec}t_{n,i}\over L_x}\right)^{1/3}. 
\label{eq:ln}
\ee
The condition
that the total drag within this layer has a negligible effect on the
reconnection layer is $\delta P_n/\rho_n\ll v_A^2$ or $v_{x,n}\ll v_A$
(which is also the condition that the neutral ejection velocity be much
less than $v_A$).  This is equivalent to 
\be
V_{rec}^2\ll {\eta_{ambi} v_A\over L_x}.
\ee
In other words, as long as the reconnection speed is less than we would
obtain from the naive substitution of $\eta_{ambi}$ for $\eta$ in
the usual Sweet-Parker formula, we can ignore the ion-neutral drag
outside the reconnection layer and set $v_i(0_+)$ equal to $V_{rec}$
in equation (\ref{eq:vrec2}).  

A significant complication is that the momentum of the ejected neutrals may 
diffuse over a distance
greater than the value of $L_n$ given in equation (\ref{eq:ln}), thereby
increasing $L_n$ and possibly the role of the neutral drag in the
reconnection layer.  This effect will become important when
\be
{\nu_n\over L_n^2}\sim {v_{x,n}\over L_x},
\label{eq:spread}
\ee
where $\nu_n\sim c_s^2t_{n,n}$ is the neutral gas viscosity.
In this limit we need to replace equation (\ref{v2}) with 
\be
{\delta P_n\over L_x}\sim \rho_n {\nu_n\over L_n^2}v_{x,n}.
\ee
If we combine this result with equation (\ref{v1}) and the
condition that $V_{rec}L_x\sim v_{x,n}L_n$ we get 
\be
L_n\sim \left(c_s^2 t_{n,n}t_{n,i}\right)^{1/4} L_x^{1/2}.
\label{eq:ln2}
\ee
The condition that the ion-neutral drag does not dissipate a
significant fraction of the magnetic energy outside the reconnection
layer is $v_A^2\gg V_{rec}L_n/t_{n,i}$, which can be rewritten,
with the aid of equation (\ref{eq:ln2}) as
\be
v_A^2\gg V_{rec} \left({L_x^2c_s^2t_{n,n}\over t_{n,i}^3}\right)^{1/4},
\ee
or, using the definitions of $V_{max}$ and $\eta_{ambi}$,
\be
V_{rec}\ll V_{max} \left({v_A\over c_s}\right)^{1/2} 
\left({t_{n,i}\over t_{n,n}}\right)^{1/4}.
\label{eq:test}
\ee
{}For largely neutral gases
the ratio of $t_{n,i}$ to $t_{n,n}$ will be of order $x^{-1}$, and
$v_A$ is usually of order $c_s$.  Consequently,
under typical conditions in the interstellar medium equation (\ref{eq:test})
will be satisfied by a comfortable margin (cf. Table 1).

Once again we need to consider whether or not the broad, slow
outflow of neutrals from the reconnection region will result in
an accumulation of gas outside the reconnection region, along
magnetic field lines that have not yet undergone reconnection.
The appropriate diffusion coefficient is, as before, $v_A^2t_{n,i}$.
The diffusion scale is $L_n$, the thickness of the stagnation 
layer in the neutral flow.  We see from equations (\ref{uy}),
(\ref{eq:vxn}), and (\ref{eq:ln}) that this implies that the transverse
distance covered by the neutrals while diffusing into the
reconnected wedge will be 
\be
v_{x,n}{L_n^2\over\eta_{ambi}}\approx L_x\left({V_{rec}\over V_{max}}\right)^{4/3}.
\ee
Since this is always less than $L_x$ we conclude that the ejected neutrals 
can be ignored.  

This case is illustrated schematically in figure 1, with the local 
velocities of the neutrals and ions indicated by thin and thick arrows.
The material in the reconnection layer is actually moving in the
$\hat x$ direction with a speed $\sim v_A$, but since most of the
gas leaves the reconnection layer in the $\hat y$ direction as neutral
particles we have ignored the motion of the ions in that layer.

Our only remaining worry is that the transverse expulsion of neutrals
may serve to remove large numbers of ions from the reconnection
zone.  It is certainly reasonable to assume a tight coupling between
the ion and neutral transverse speeds, since the transverse shear
rate, $v_{x,n}/L_x$ is much smaller than the ion coupling rate,
$(\rho_n/\rho_i)t_{n,i}^{-1}$.  However, for typical ISM parameters
we also have $\tau_{recomb}^{-1}> v_A/L_x>v_{x,n}/L_x$, so the
ion fraction in the gas is maintained even while individual
ions are expelled.  We also note that even if we ignored
recombination the equations of continuity for ions and neutrals
combine to give  
\be
v_i \partial_y \ln \rho_i - v_n\partial_y\ln\rho_n+\nabla (v_i-v_n)=0.
\ee
Within the outer layer of thickness $L_n$ the neutral particles
come to a halt and $v_i$ remains close to $V_{rec}$.  Consequently
the density scale height for the ions is not less than $L_n$ and
some large fraction of the ions will reach the resistive layer,
despite transverse losses.

We conclude that
\be
V_{rec}=
\left({2\eta_\infty\over\tau_{recomb,\infty}}\right)^{1/2}
\left({T_\infty\over T(0)}\right)^{3/2}
\left(1+{1\over\beta x}\right)~~~,
\label{eq:vrec3}
\ee
where $\beta$ is the ratio of the gas pressure to the
magnetic pressure.
The width of the reconnection layer is
\be
\Delta={\eta(0)\over V_{rec}}=(\eta_\infty\tau_{recomb,\infty})^{1/2}
\left(1+{1\over\beta x}\right)^{-1}~~~.
\label{Ly}
\ee

The principle source of ion heating in the reconnection layer is
ohmic heating due to the dissipation of the magnetic field.  Cooling
can take place either through radiative losses or through the transfer
of energy to the neutral gas.  The neutrals will, in turn, shed their
excess thermal energy through radiative losses and/or by escaping from
the reconnection layer and diluting the extra heat over a large volume.
Since neutral mean free path is typically much larger than $\Delta$, and 
since $c_s\gg \Delta/t_{n,i}$, the ion cooling rate will be controlled by
the ion-neutral collision rate.  In other words,
\be
{B_\infty^2\over 8\pi}{\eta(0)\over\Delta^2}\approx {3\over2}2n_ik_BT(0){2\over t_{i,n}},
\label{eq:hot}
\ee
where we have assumed that $T(0)\gg T_{\infty}$, since otherwise
the correction is of no interest, and
used $B_\infty$ since we have already shown that 
$B_\infty\sim B(0_+)$ for reconnection in the interstellar medium.
We have also assumed that the electrons and ion share the same temperature
in the reconnection layer.
Given that the charged particle pressure in the ionization zone
balances the external magnetic pressure, equation (\ref{eq:hot})
implies
\be
{V_{rec}^2\over \eta(0)}\approx {3\over t_{i,n}}.
\ee
We can rewrite this with the aid of equation (\ref{eq:vrec3})
as
\be
\left(1+{1\over\beta x}\right)^2\approx 
\left({3\tau_{recomb,\infty}\over t_{i,n}}\right)
{\phi(T_{\infty})\over \phi(T(0))} 
\left({T(0)\over T_\infty}\right)^{5/2}.
\ee
This implies that when reconnection layer heating is important we
should substitute the expression
\be
\left({T_\infty\over T(0)}\right)^{3/2}\approx 31
\left(1+{1\over\beta x}\right)^{-6/5}x^{-3/5}\phi(T(0))^{-3/5}T_\infty^{0.3},
\label{eq:rlh}
\ee
into equation (\ref{eq:vrec3}).  We see at once that $T(0)>T_{\infty}$,
so that this correction is appropriate, only in cold regions with
$x\ll 10^{-3}$ (assuming $\beta$ is of order unity).  As we reach
this limit we go from $V_{rec}\propto x^{-0.5}$ to $V_{rec}\propto x^{0.1}$.

It is important to remember that for strictly one dimensional formulations
of this problem there is no natural choice for $L_n$ and the actual
reconnection speed will remain sensitive to the computational box size
and/or the boundary conditions.  Although $L_x$ does not appear in
equation (\ref{eq:vrec3}) it is still present implicitly as part of the
constraints that make this a well-posed problem.  Purely one dimensional
calculations will not capture all the relevant physics, and may result
in a wide variety of estimates for $V_{rec}$ and $\Delta$.

Equation (\ref{eq:vrec3}) constitutes the main result of this paper,
and may be regarded as a generalization of the Sweet-Parker reconnection
rate to an ambipolar medium with a long recombination time.  The
major effect we have neglected here is additional ionization within
the reconnection layer caused by the reconnection process.
We have included the effect of local heating on the 
resistivity, gas density, and recombination rate. 
We note that even taking $T(0)=T_\infty$
we can show that reconnection in this model
is rather slow.

We should note that 
equation (\ref{eq:vrec3}) is correct only if $\tau_{recomb}$ is much less
than the shearing time $L_x/v_A$. In the opposite limit, ions
will escape from the sides of reconnection zone rather than
through recombination.   In this case we are almost back in
the regime described in section 2.1.  However, there are two important
physical effects which have to be considered.  First, we have
already seen that the partial decoupling of ions and neutrals 
implies a large concentration of ions in the reconnection layer itself.
Consequently the rate of ion ejection is enhanced by the
ratio $n_i(0)/n_{i,\infty}$.  The ion conservation equation
becomes
\be
\Delta v_{eject} n_i(0)=L_x V_{rec} n_{i,\infty}.
\label{eq:mconsp}
\ee
Second, the neutrals will diffuse out the reconnection layer, spreading
the transverse momentum and creating an additional drag on the ions.
Since these effects work in opposite directions, it's not immediately
obvious whether the presence of neutrals increases or decreases the
reconnection speed when $\tau_{recomb}$ is very large.

Balancing the ion pressure gradient in the $\hat x$ direction with
the neutral drag, we get
\be
{\delta P_i\over L_x}\approx \rho_i(0) {\Delta v_x\over t_{i,n}}.
\ee
The transfer of momentum to the neutrals is then balanced by the 
viscous drag on the neutrals, so that
\be
\rho_i(0) {\Delta v_x\over t_{i,n}}\Delta \approx \rho_n v_{eject}
{\nu_n\over L_n},
\ee
where we have assumed that $\Delta v_x$, the difference between
the ion and neutral particle velocities in the $\hat x$ direction,
is small compared to $v_{eject}$.  The width of the spread of
$\hat x$ momentum is given by equation (\ref{eq:spread}) so
we conclude that
\be
v_A^2\Delta\approx \left(v_{eject}^3 L_x\nu_n\right)^{1/2}.
\label{eq:momconsp}
\ee
Remembering that $\Delta\approx \eta(0)/V_{rec}$ we can combine
equations (\ref{eq:mconsp}) and (\ref{eq:momconsp}) to obtain
\be
V_{rec}\approx\left({v_A\eta(0)\over L_x}\right)^{1/2}
\left({\eta(0)\over\nu_n}\right)^{1/8}
\left({n_i(0)\over n_{i,\infty}}\right)^{3/8}.
\ee
Consequently, we can estimate the corrected Sweet-Parker reconnection
velocity as
\be
V_{CSP}\approx\left({v_A\eta_\infty\over L_x}\right)^{1/2}
\left({\eta_\infty\over\nu_n}\right)^{1/8}
\left({T_\infty\over T(0)}\right)^{21/16}
\left(1+{1\over x\beta}\right)^{3/8}.
\label{esp}
\ee
In spite of the fact that typically $\nu_n\gg\eta_\infty$, this
estimate will usually be larger than the standard SP reconnection
speed for largely neutral, magnetized gas.

{}Finally, we note that our estimates are based on the assumption 
that the reconnection layer is in a steady state, meaning that
the local pressure excess is comparable to the magnetic field
pressure and that the accumulation of ions in the reconnection
layer has reached a steady state.  If the magnetic field lines
are being compressed by turbulent forces then the former condition will
be reached in about one eddy turnover time, or, assuming 
$v_A\sim v_{turb}$,
about $L_x/v_A$.  Otherwise we should replace the eddy turn over
time with the large scale time dynamical time scale.
At earlier times the confining pressure will
be less than our estimates, and the reconnection rate slower.
On the other hand, filling up the reconnection layer with ions
will take a time $\sim (\Delta/V_{rec})(n_i(0)/n_{i,\infty})$
and at earlier times the reconnection layer will be compressed
to a smaller width, with a consequent increase in the reconnection
rate.  We can rewrite this time scale using equation (\ref{eq:vrec3}).
We obtain
\be
{\Delta\over V_{rec}}{n_i(0)\over n_{i,\infty}}\approx
{\eta(0)\over V_{rec}^2}{n_i(0)\over n_{i,\infty}}\approx
\tau_{recomb,\infty}\left({T(0)\over T_\infty}\right)^{1/2}
\left(1+{1\over x\beta}\right)^{-1}.
\ee
This will usually be somewhat less than an eddy turnover time in
the interstellar medium.  We conclude that
the early phases of reconnection are apt to be characterized by
weak compression and slow reconnection, and in any case that 
the reconnection of magnetic flux is unlikely to proceed much
more rapidly in the early phases of reconnection.

\section{Phases of the ISM and reconnection}

Does ambipolar diffusion actually lead to a dramatic enhancement
of reconnection rates in the ISM?  In Table~1 we give our
results for idealized partially ionized phases of the ISM. 
We include, for reference, $V_{max}$, the reconnection velocity
when $\tau_{recomb}\rightarrow0$, and $V_{SP}$, the Sweet-Parker
reconnection speed without corrections for natural drag or ion
density enhancement.  More realistically, we also show $V_{rec}$,
in which the reconnection speed is limited by the recombination
rate, and $V_{CSP}$, the corrected Sweet-Parker reconnection speed. 
{}For molecular hydrogen clouds the reconnection layer heating is
important in $V_{rec}$,  and has been included using equation
(\ref{eq:rlh}).  The reconnection
velocities $V_{SP}$ and $V_{CSP}$ will depend on the assumed
values of $L_x$. Here we take $L_x=1$~pc so that both these
speeds scale as (1~pc$/L_x$)$^{1/2}$.  We also make use of the evidence 
that for all the phases shown in Table~1 the Alfv\'en velocity 
$v_A$ is of the order of a couple of kilometers per second. 
Here we assume that it is $10^{5}$~cm s$^{-1}$.  Finally, each
ISM phase we give $nf$, the mass fraction of that phase.
A discussion of the values we use here is contained in
Draine and Lazarian (1997).  We note that 
an additional 22\% is contained in ionized gas.

There are several important points to be noted in connection
with Table 1.  First for all the phases presented in Table~1 
it is the ``recombination limited'' reconnection rate 
(see  Eq~(\ref{eq:vrec3})) that determines the reconnection speed
on scales of $\sim$ 1~pc. Apart from diffuse warm HI, where reconnection
is slow anyhow, the corrected Sweet-Parker reconnection speed given by
equation~(\ref{esp}) is at least one order of magnitude slower than
$V_{rec}$.  This situation changes only at much
smaller scales where outflows from reconnection regions can carry
out most of the ions faster than recombination can eliminate them.

Second, in all phases of the ISM, $V_{max}$ is clearly a gross
exaggeration of the degree to which ambipolar diffusion can
enhance reconnection.  Nevertheless, $V_{rec}$ is very much
larger than $V_{CSP}$ for all the cold phases of the ISM.
The latter is, in turn, considerably larger than $V_{SP}$.
In both cases, this increase is due to the compression of
ions in the reconnection layer.  The conclusion that
$V_{rec}>V_{CSP}$ is due to the double contribution of the
ion density enhancement in the former case, that is, not only
does it contribute directly by enhancing the efficiency of
ion loss relative to the incoming flux of ions, but it also
contributes by raising the recombination rate.  This explains
why the warm neutral gas shows only a small increase in 
its reconnection speed.  It also suggests that weak
magnetic fields will have much slower reconnection speeds. 

Third, $V_{rec}$ is always much smaller than $v_A$.  The 
only case in which it exceeds $10^{-3}\times v_A$ is that of
dense molecular gas.  Moreover, this phase has a modest mass
fraction of about 3\%.
While this result implies that reconnection should be efficient
enough to remove magnetic flux from star forming regions, and
disconnect the internal magnetic field from its environment, it
does not explain how reconnection could be fast enough to sustain the
galactic dynamo.

{}Finally, these calculations are based on the assumption that the plasma 
can be treated as a fluid.  Given the low collision rates in the ISM,
this assumption requires, at a minimum, that all scales of interest
be much greater than the ion Larmor radius.  In particular, we require
that $\Delta\gg R_{L,ion}$ or
\be
1\gg {V_{rec}\over\eta(0)}{V_{th,ion}m_{ion}c\over eB}
\approx \left({2\over\eta_{\infty}\tau_{recomb,\infty}}\right)^{1/2}
\left(1+{1\over\beta x}\right){V_{th,ion}\over v_A}
{cx^{1/2}\over\omega_{p,ion}},
\ee
where $\omega_{p,ion}$ is the ion plasma frequency.
If we evaluate this expression for the idealized ISM phases shown
in Table 1 we find that it rises from a low of about $3\times10^{-2}$
for diffuse cold $HI$ to about $0.13$ for diffuse molecular gas.
There is a weak dependence on reconnection layer heating which
increases the ratio for molecular gas.  As before, we have calculated
the local heating using equation (\ref{eq:rlh}).  For dense molecular
gas the ion Larmor radius is roughly the same as the reconnection
layer thickness.  This suggests that it would be appropriate
to reconsider this case without the use of the fluid approximation.
Such a calculation is beyond the scope of this paper.  We merely
note that our results for this case should be treated with caution.
However, it seems unlikely that the reconnection rates will decrease 
significantly when we allow for the finite Larmor radius of the ions. 
If $r_{L,ion}\gg \Delta$ then the ions do not feel the magnetic fields in 
the reconnection layer. Nevertheless, they will be confined to this layer
by the electric field of the electrons, which are in turn confined by
the magnetic pressure exerted on the current-carrying component of the plasma. 
It follows that the electrons and ions will be moving together as in the hydrodynamic
approximation. Possible deviations from this approximation in the form of
instabilities, if they take place, will only enhance magnetic diffusivity 
and reconnection.

\section{Discussion and Conclusions}

In this paper we have examined the role of neutral particles
on the structure of reconnection regions in the ISM.  In so doing,
we have assumed that the basic geometry posited by Sweet and Parker 
accurately describes reconnection in three dimensions.  We find
that ambipolar diffusion can give rise to much faster reconnection
speeds than those obtained by the naive use of the usual Sweet-Parker
formula, but these speeds are still very small compared 
to the local Alfv\'en speed.  In addition, most of the increase is
due to the compression of ions in the reconnection layer, and 
would not occur in a fluid where the ion and magnetic pressures
were comparable.
 
How can we interpret these results?  Obviously, there are 
physical effects which we have ignored here, but which might
affect the geometry or the physics of reconnection.  Assuming
our results are a realistic description of reconnection in the
ISM we can examine their implications for the structure of the
magnetic field in the ISM.  We will address the latter issue
first, and then briefly take up the prospect of obtaining faster
reconnection speeds under realistic circumstances.

What have we learned about flux tube formation?  In a previous
paper (Vishniac 1995) one of us showed that flux tubes should be 
a natural feature of turbulent high beta plasmas.  This is consistent
with observations of the solar photosphere, but not otherwise directly
testable.  In parts of the interstellar medium
the effective plasma beta can be large, if one accounts not only 
for thermal plasma pressure, but for the pressure from compressional
shocks and cosmic rays (Lazarian \& Vishniac 1996).
Therefore, it seems possible that flux tubes could form in the 
interstellar medium.  The basic mechanism involves the
process of turbulent pumping, in which flux tubes are stretched
by the surrounding turbulence, and then twisted to produce
close loops of flux which self-destruct and release their
entrained ions into the surrounding medium.  This assumes that
reconnection is efficient.

However, our results indicate that ambipolar diffusion
acts against the flux tube formation process.  In addition to the
small reconnection speeds we have obtained, ambipolar diffusion
makes the tubes leaky since it allows neutrals to cross magnetic 
field lines, diffuse into the flux tubes, and subsequently fill
them up with newly produced ions.  When recombination times
are short we can show that  flux tubes can still be formed. 
In this case ambipolar diffusivity works in 
exactly the same way that Ohmic resistivity works in a completely
ionized plasma. The trade-off between more rapid ambipolar reconnection
and ambipolar
leakage results in a rather weak dependence of turbulent pumping
efficiency on the ambipolar diffusivity. The situation changes
dramatically if the recombination times are large, as they are
in the ISM. In this case ambipolar reconnection alone cannot provide 
the necessary rates of turbulent pumping.

We conclude that flux tubes will not form in the ISM unless there
are ways to obtain dramatically higher reconnection speeds.  The
only possible exceptions are the dense molecular parts of the ISM.
On the other hand, we note that the best evidence for the
existence of flux tubes comes from the solar photosphere, where the high
collision rates imply that ambipolar diffusion is only slightly
more effective than ohmic diffusion.  In addition, the best
evidence for high reconnection speeds comes from parts of the
solar chromosphere and corona where the neutral fraction is negligible. 
Evidently there must be circumstances where reconnection is much
faster than our estimates and these may allow the formation of
flux tubes in the ISM.

What do our results say about the galactic dynamo?  Our estimates for
$V_{CSP}$ depend on $L_x$, and reconnection might occur over
small distances at speeds higher than those we have given here. 
On the other hand, we have already assumed $L_x$ is of order
a parsec, and the large scale eddies in the ISM may be as much
as two orders of magnitude larger.  Moreover, it is precisely
these large scale motions which are usually thought to
drive the galactic dynamo.  Finally, we note that $V_{rec}$ is
insensitive to $L_x$.  While it is difficult to be precise about
the reconnection speeds necessary to allow turbulent diffusion and
the existence of a large scale dynamo, the usual assumption is
that these speeds must be in the range 0.01 to 0.1 times
$v_A$ (see Ruzmaikin, Sokoloff \& Shukurov 1988) 
or the magnetic field will become tangled on the smallest
scales and show little evidence for large scale organization.
Our largest estimates for $v_A$ falls just outside this range
and applies to only a few percent of the interstellar gas.
Once again our results are disappointing in this regard and suggest
the necessity for some way to obtain much larger reconnection speeds.

We could consider circumstances where the gas density is much larger,
for example, a cool stellar atmosphere.  The gas in the solar
photosphere is largely neutral and dense enough that recombination
will be very fast.  On the other hand, under these conditions $t_{i,n}$
is also very much shorter, and ambipolar diffusion
is not very much faster than ohmic diffusion.  For example, if we take 
$n_n\sim 10^{10}$,
and $B\sim 1$~gauss then $\eta_{ambi}\sim 10^{11}$, about $10^2$ larger
than the ohmic diffusivity in a cool stellar atmosphere.  This
suggests a reconnection speed an order of magnitude larger than the
Sweet-Parker estimate, but still quite small compared to turbulent
diffusivity in a giant stellar atmosphere.

Are there additional physical effects that might affect our results?
We start by considering cosmic rays.  Cosmic rays are an essential 
component of the interstellar medium and their pressure is roughly 
of the same order as the pressure of the magnetic field in the ISM, suggesting 
that they may play an important role in all plasma processes in the ISM, 
including reconnection.  Our discussion here will follow 
Longair (1994).

Cosmic rays diffuse through magnetic field and the magnetic
fluctuations on the scale of the Larmor radius are the most effective 
source of scattering. Therefore for our order of magnitude estimates we may
consider that the time for scattering of cosmic rays is of the order
$t\sim N/\omega_L$, where $\omega_L$ is the Larmor frequency and $N$
can be defined as the number of orbits required to substantially alter the initial
motion of the charged particle. In other words, if in the course of an
individual scattering event the particle deviates over an angle
$\Phi\sim B_{rms}/B_{reg}$, where $B_{rms}$ and $B_{reg}$ are the
rms and regular components of the magnetic field at the place of scattering,
then $N$ satisfies the relation: $N^{1/2}\Phi\sim 1$, where the 
exponent $1/2$ appears above due to the random walk nature
of the process.
As a result the diffusion coefficient for a charged particle
with elementary charge $e$ and energy $E$  can be
defined as:
\be
\eta_{cr}\sim \frac{r_L^2}{t}\sim \frac{cE}{eB}\Phi^{2}~~~.
\ee
In the zone of reconnection cosmic rays are streaming along the
magnetic field lines and perturb magnetic field lines through nonlinear 
interactions.  In these circumstances $\Phi$ does 
not differ much from unity. As a result values of $\eta_{cr}$ can be as large 
as $10^{20}$ cm$^2$ s$^{-1}$ for the Mev charged particles. If we assume
that the background gas density is negligible, as it may be in the
galactic halo, then the Alfv\'{e}n
speed approaches that of light, and we obtain high reconnection rates
(\cite{P92}).

In the galactic disk, where the density of gas is not negligible,
the situation is quite different.  In general the Larmor radius for
cosmic ray particles will be much larger than the reconnection layer
width.  Consequently, cosmic rays cross the reconnection zone in
a single orbit.  Moreover, neutral particles damp the Alfv\'{e}nic waves,
thereby reducing the coupling between cosmic ray streaming and
perturbations to the local magnetic field.  Finally, the reduced
magnetic field in the reconnection layer decreases the coupling between
the plasma and the cosmic rays.  In the end, the only obvious role
for the cosmic rays is that they increase the effective pressure
outside the reconnection zone, and consequently help compress the
gas in the reconnection layer. 
This changes the relation between the mass influx and mass efflux so
that: 
\be
\rho_1 L_x u_x=\rho_2 L_y u_y~~~,
\ee
where 
\be
\frac{\rho_2}{\rho_1}=\frac{P_{gas}}{P_{gas}+P_{cr}}~~~.
\ee
Above $P_{gas}$ and $P_{cr}$ are the gaseous and cosmic ray pressure
respectively.  This increases the reconnection speed by the square
root of the density ratio.  Since the cosmic ray pressure is roughly
the same order as the gas pressure in the interstellar medium, this
implies that cosmic rays change the reconnection
rate just by a factor of order unity.

How robust is our assumption concerning the basic reconnection geometry?
It is well known that the kind of narrow current sheet we are positing
as the zone of reconnection is dynamically unstable to tearing modes
(\cite{FKR63}).  This instability persists for reconnection in
largely neutral environments, as noted by Brandenburg and Zweibel (1995).
Turbulent mixing of the current sheets will promote reconnection
by enhancing the transport of fresh magnetic flux into the reconnection
zone and by mixing the accumulating ions outward.  In fact, at least
one author (\cite{S88}) has proposed that this will lead to 
reconnection speeds of order $(\delta B/B)v_A$ and shown that $\delta B/B$
will be be of order unity in saturated turbulence.  Although some
enhancement of reconnection speeds seems inevitable, there are grounds
for doubting that this will dramatically change our results.
We will defer a detailed discussion until our next paper, but preliminary
estimates indicate that allowing for the effects of three dimensional
nonlinear tearing modes in
an ionized medium will increase the reconnection speed over the Sweet-Parker
estimate by only a factor of the magnetic Reynolds number to the one sixth
power.

{}Finally, and on a more hopeful note, the existence of a third dimension may allow the 
interpenetration of magnetic field lines for favorable geometries,
i.e. an interchange instability, which will facilitate reconnection
by dramatically increasing the surface area of the reconnection layer.
(cf. Uchida \& Sakurai\ 1977).  We will explore these ideas in a 
subsequent paper.

\acknowledgements

We are grateful to E. Zweibel, B. Draine and R. Kulsrud for a 
series of helpful discussions.  This work was supported in
part by  NASA grants NAG5-2858 (AL), NAG5-2773 (ETV), and
NSF grant AST-9318185 (ETV).  ETV
is grateful for the hospitality of MIT and
the CfA during the completion of this work.

\eject

\begin{deluxetable}{lccccc}
\tablewidth{40pc}
\tablecaption{Reconnection speeds for idealized phases of the ISM}
\tablehead{
\colhead{ISM phase}      & \colhead{$\tau_{recomb}$} &
\colhead{$V_{max}$}      & \colhead{$V_{rec}$}       &
\colhead{$V_{SP}$}         & \colhead{$V_{CSP}$}}
\startdata
diffuse cold HI& $4.0\times 10^{12}$ & $2.3\times 10^3$    & $7.8$ 
                               & $9.3\times 10^{-3}$ & $1.4\times10^{-1}$\nl
$n=30$ cm$^{-3}$, $x=10^{-3}$& & & & \nl  
$T=100$~K, $nf\approx 0.15$ & & & & \nl
\hline
diffuse warm HI& $3.6\times 10^{13}$ & $1.9\times 10^3$    &
$4.1\times 10^{-4}$                  & $4.3\times 10^{-4}$ & 
$3.0\times 10^{-4}$ \nl
$n=0.3$ cm$^{-3}$, $x=0.15$ & & & & \nl         
$T=6000$~K, $nf\approx 0.4$ & & & & \nl
\hline
diffuse H$_2$  & $1.6\times 10^{13}$ & $5.6\times 10^3$    & $(38)$ 
                                     & $1.4\times 10^{-2}$ & 
$0.56$ \nl
$n=50$ cm$^{-3}$, $x=10^{-4}$ & & & & \nl 
$T=60$~K, $nf\approx 0.2$ & & & & \nl
\hline
dense H$_2$    & $4.5\times 10^{12}$ & $3.9\times 10^3$    & 
$(220) $                   & $3.1\times 10^{-2}$ & 4.0\nl
$n=10^3$ cm$^{-3}$, $x=10^{-5}$ & & & & \nl
$T=20$~K, $nf\approx 0.03$ & & & & \nl
\hline
\enddata
\tablecomments{
All quantities are given in cgs units.
$\tau_{recomb}$ is the recombination time outside the reconnection layer;
$V_{max}$ follows from infinitely fast recombination 
(Eq.~(\ref{uy}));
$V_{rec}$ allows for a realistic recombination rate
(Eq.~(\ref{eq:vrec3})); $V_{SP}$ is the Sweet-Parker reconnection rate 
(Eq.~(\ref{eq:sp1})); and $V_{CSP}$ is the corrected Sweet-Parker 
reconnection
speed given by Eq.~(\ref{esp}). $V_{rec}$ for H$_2$ is
placed in brackets as it allows for heating of the reconnection layer.}
\end{deluxetable}
\clearpage
\begin{figure}
\begin{picture}(441,216)
\includegraphics{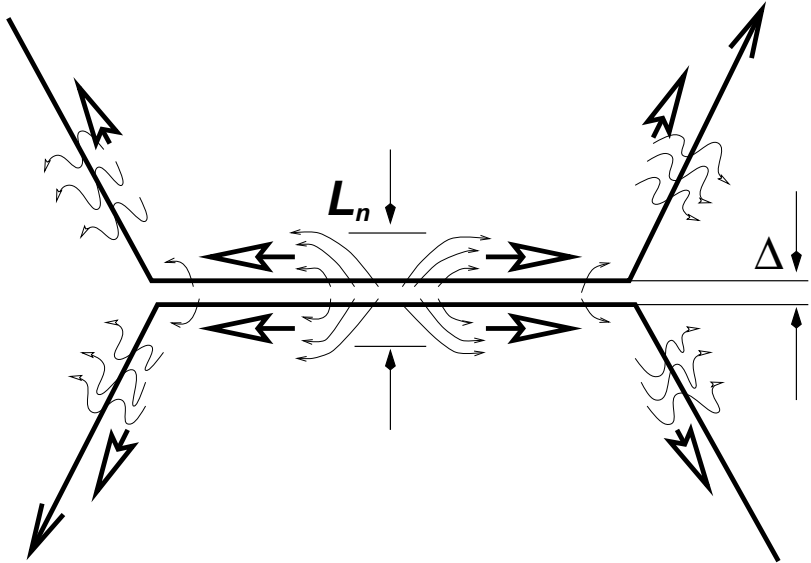}
\end{picture}
\figcaption{A schematic of a reconnection region.  Magnetic reconnection
is occuring in a layer of width $\Delta$.  Gas velocities are marked
with thick lines for ions and thin lines for neutrals.  The gas
outflow is confined  to a region of width $L_n$. The $\hat x$
axis is horizontal and the $\hat y$ axis is vertical.}
\end{figure}
\end{document}